\documentclass[showpacs,preprintnumbers, amsmath,amssymb,superscriptaddress,preprint]{revtex4}
\usepackage{graphicx}
\draft
\begin{document}
\newcommand {\be}{\begin{equation}}
\newcommand {\ee}{\end{equation}}
\newcommand {\bea}{\begin{eqnarray}}
\newcommand {\eea}{\end{eqnarray}}
\newcommand {\nn}{\nonumber}


\title{Adaptive design of nano-scale dielectric structures for photonics}
\author{Yu Chen}
\address{Department of Electrical Engineering, University of Southern
California, Los Angeles, CA 90089-2533}

\author{Rong Yu}
\address{Department of Physics and Astronomy, University of Southern
California, Los Angeles, CA 90089-0484  \\ \textrm{(Received}}

\author{Weifei Li}
\address{Department of Physics and Astronomy, University of Southern
California, Los Angeles, CA 90089-0484  \\ \textrm{(Received}}

\author{Omid Nohadani}
\address{Department of Physics and Astronomy, University of Southern
California, Los Angeles, CA 90089-0484  \\ \textrm{(Received}}

\author{Stephan Haas}
\address{Department of Physics and Astronomy, University of Southern
California, Los Angeles, CA 90089-0484  \\ \textrm{(Received}}

\author{A.F.J. Levi}
\address{Department of Electrical Engineering, University of Southern
California, Los Angeles, CA 90089-2533}
\address{Department of Physics and Astronomy, University of Southern
California, Los Angeles, CA 90089-0484 \\ \textrm{(Received}}
\textrm{\textit{}}

\begin{abstract}
\setlength{\baselineskip}{8mm}
Using adaptive algorithms, 
the design of nano-scale dielectric structures for photonic applications
is explored. 
Widths of dielectric layers in a linear array are adjusted to match target 
responses of optical transmission as a function of energy. Two complementary 
approaches are discussed. The first approach uses adaptive local random 
updates and progressively
adjusts individual dielectric layer widths. The second approach is based on 
global updating functions in which large subgroups of layers are adjusted
simultaneously. Both schemes are applied to obtain specific target responses
of the transmission function within selected energy windows, such as 
discontinuous cut-off or power-law decay filters close to a photonic band edge. 
These adaptive
algorithms are found to be
effective tools in the custom design of nano-scale photonic 
dielectric structures. 

\end{abstract}
\maketitle

\newpage
\section{Introduction}
\setlength{\baselineskip}{8mm}
In recent years, several types of optical filters, superprisms, and distributed mirrors
have been suggested which make use of photon propagation in nano-scale 
dielectric structures.\cite{kosaka,gralak,baba}
While traditional approaches in the design of these devices are based on
spatially symmetric arrangements of
dielectrics,\cite{ohtaka} this study explores the merits
of intentionally breaking translational symmetry to better realize 
desired target response profiles, such as transmission or reflection
as a function of
incident light energy.\cite{skorobogatiy} 
It is broken symmetry that enables useful photonic functions.
In this paper, two prototype algorithms are discussed which use
either local or global progressive updates of dielectric
layer widths to match 
target optical transmission functions, such as a cut-off
or a power-law decay filter within a given frequency window. 

To illustrate our approach, we
focus on the physical problem of one-dimensional arrays 
of dielectric optical  ``wells" and ``barriers"
with alternating refractive indices $n_1 = 1.0$
and $n_2 = 1.5$, respectively. Monochromatic light with energy $E = h \nu$ is incident 
from the left hand side, and transmission is detected on the right end
of the array. The propagation matrix method, keeping track of the 
boundary conditions on the reflected and transmitted components of the 
wave function at each 
individual barrier, is applied to obtain the total transmission
coefficient of the array as a 
function of the photon energy. For the case of a symmetric array
with constant barrier width $b(x) = b_0 = 0.5 \; \mu m$, shown 
in Fig. 1(a), this 
leads to a typical response profile, Fig. 1(b),
containing bound state resonances
at low energies, and a photonic band edge at $E = 0.428\; eV$.   
The resonances are due to the finite size of the barrier structure.
On the other hand, the slightly randomized array with
$b(x) = b_0 \pm \Delta b(x)$ in Fig. 1(c),
shows a clear overall reduction of transmission, Fig. 1(d),
while still displaying 
remnant features of the symmetric case, such as the band gap. 
It is our objective to utilize
such intentional translational distortions of symmetric barrier arrays    
to match target optical response functions, i.e. reflection
and transmission, in a given energy window.
The optical response of a system with N barrier pairs is determined
by the contribution of N-1 barrier poles. Desired filter functions 
can then be generated over a finite range of energy by adjusting the 
contribution of each pole. 

\section{Local updates by guided random walk}
The first type of adaptive algorithm to achieve this task is based on 
local random
updates of individual barrier widths. These updates are accepted if the 
resulting transmission profile matches better the target function than 
the previous one. The basic steps of the algorithm are:
\begin{enumerate}
{\item Choose target function $T(E)_{target}$ and energy window
$E \in [E_{min},E_{max}]$, e.g. cut-off function $T(E)_{target} 
\equiv \theta (E - E_c)$  with $E_c \in [E_{min},E_{max}]$.}
{\item Generate initial barrier array by setting 
length and refractive index of each barrier, for example in a 
spatially symmetric fashion.}
{\item Determine $T(E)$ for initial array, and 
compute its deviation from the target function by evaluating 
$\Delta = \int_{E_{min}}^{E_{max}} dE [T(E) - T(E)_{target}]^2$.}
{\item Perform a trial random 
update (change of width) of one barrier (or sets of barriers),
and determine $\Delta$ 
for the following configuration. Additional physical constraints, such as inversion 
symmetry about the array center, can be enforced in this step.}
{\item Accept the update if $\Delta$ has decreased
with respect to the previous configuration.} 
{\item Repeat random updates of all barriers until acceptable convergence 
has been reached.} 
\end{enumerate}

In principle, this annealing approach can be further improved by 
(i) implementing a Metropolis criterion that avoids local minima in the 
phase space of barrier widths, and (ii) choosing more updates close to the 
array boundaries, which affect $T(E)$ the most, than in the vicinity
of the center of the structure. 
In practice, however, the convergence 
of this algorithm proves to be sufficiently fast. To illustrate this 
point, let us examine an array of 30+2+30=62 barriers with alternating 
refractive
indices $n_1 = 1.0$ and $n_2 = 1.5$. 
Perfect inversion symmetry of the barrier widths about
the array center is enforced, and the four central barriers are kept
unchanged at their initial duty cycles. The reason for this additional 
constraint is the physical motivation to create an array of adjustable optical 
barriers that smoothly modulates an incoming free light wave into a crystal 
Bloch wave through an intermediate layer. 

For the symmetric case with a dielectric barrier pair width
$d = 1 \; \mu m$ and an individual barrier width of $b_0= 0.5 \; \mu m$,
this array 
exhibits a band edge at $E_c = 0.43251\; eV$. This is a natural point in 
energy to center a target cut-off filter function of the shape 
$T(E)_{target} \equiv \theta (E - E_c)$ within a given energy window,
which we choose as
$E \in [0.35\; eV,0.45\; eV]$. The result of 150 successful updates is 
displayed in Fig. 2. 

In Fig. 2(a) the transmission function for the translationally
symmetric array with 
equal barrier widths $b_0$ is shown. It displays characteristic
resonances that increase close to the band edge $E_c$. Comparing this 
response with $T(E)$ after 
the application of the adaptive local random update algorithm, shown in 
Fig. 2(b), one observes that in
both cases the band edge is the dominant feature. However, after convergence
to the target filter function, the oscillations in $T(E)$ are largely
suppressed 
within the chosen energy window. The deviation of $T(E)$
from $T(E)_{target}$ is plotted in Fig. 2(c)
as a function of accepted updates. 
From a fit to the form $\Delta_j =
A \exp{(-j/\chi_j)}$, where $j$ is the label of 
successful updates, it is found that this algorithm converges 
exponentially fast on a scale of $\chi_j \approx 27$ updates.    
After approximately 100 successful
updates this function is essentially flat at $\Delta_{\infty}
\simeq 0.004 \Delta_0$, and 
convergence has been reached. 
The strength function of the final configuration of barrier widths is
displayed in Fig. 2(d). 
This quantity is defined by $s(x) = 2 w(x)/d$, where $w(x)$ is the barrier
width at position $x$, and $d = 1 \; \mu m$ is the dielectric barrier pair 
width. From this last figure it is obvious that the final 
configuration does not have the simplest spatial symmetry,
although 
inversion symmetry about the array center has been enforced. The
barrier widths 
in the array center remain almost unchanged, whereas 
they decay rapidly towards the array extremities. Therefore, adjustments
in these boundary regions prove to be most effective in achieving a 
target optical response.   
Moreover, from this 
example of multiple independently adjustable barriers (with wells)\cite{footnote1}
it is evident that there
are several sets of possible 
solutions $\{ w(x), x=1,N \} $ for a given finite tolerance $\Delta $.  
This number of local minima can be reduced by enforcing lower
symmetries, such as the inversion about the array center that was 
used in the above example. 
However, for more complicated target functions such symmetries may not 
exist or are not obvious. Furthermore, if fewer 
symmetries are enforced by hand, there are more adjustable parameters, 
and the deviation
from the target function can be reduced more efficiently.
Finally, often practical applications 
do not require tolerances below the $1\%$-level that is reached 
by this local random update algorithm.

\section{Global updates using physical constraints}
Let us next turn to an alternative 
approach to the adaptive design of barrier arrays that it is based on global 
updates of the strength function $s(x)$. More specifically, the 
strength function is expanded in a set of basis functions $f_i(x)$, 
$
s(x) = \sum_{i=0}^n a_i f_i(x) 
$,
which are subject to the constraints 
(i) $s(x)=s'(x)=0$ at the system boundaries, (ii) $s(L)=1$ and 
$s'(L) = 0$ at or close to
the array center, and (iii) $s(x)$ 
is forced to be inversion-symmetric about the array center. 
The first two
constraints are used to determine the lowest four coefficients in the 
expansion of $s(x)$, i.e. $a_0$ up to $a_3$. The remaining coefficients
$a_4$ up to $a_n$ are then determined numerically, optimizing the 
overlap of the transmission in a given energy window with the target 
function by tuning $\Delta (a_i ) $ with $i = 4, ..., n$. This approach is
only partially numerical.
The algorithmic steps can be summarized  as follows: 
\begin{enumerate}
{\item Choose target function $T(E)_{target}$ and energy window
$[E_{min},E_{max}]$.
}
{\item Generate initial barrier array by setting 
length and refractive index of each barrier.
}
{\item Determine $T(E)$ for initial array, and 
compute its deviation from the target function by evaluating 
$\Delta = \int_{E_{min}}^{E_{max}} dE [T(E) - T(E)_{target}]^2$.}
{\item Expand strength function in a finite basis, e.g. polynomials, 
and determine lowest four coefficients from boundary conditions.}
{\item Find minimum of strength function as a function of higher
coefficients by numerically minimizing $\Delta (a_4, ...., a_n)$.
}
\end{enumerate}

Assuming that the expansion is about a global minimum of $\Delta (a_n)$,
the minimization can be performed sequentially, i.e. by first finding 
the optimum value of $a^o_4$ for the expansion to $n = 4$, and then 
optimizing $\Delta (a_0, ... ,a_5)$  for fixed $a^o_4$, etc. Furthermore, this
scheme is improved by adjusting the precise position of the cut-off
energy in the target function according to the location of the band edge
after each optimization step.  
In Fig. 3 results for the cut-off target function are shown. The strength 
function is expanded in a polynomial basis up to $x^5$, and boundary 
conditions are applied to determine the lowest expansion coefficients. The
initial parameters are chosen to be identical to the previous discussion 
of the local random update
algorithm. 


In Fig. 3(a) the transmission function of the globally adjusted barrier array 
is shown. Compared with the result of the local random update algorithm,
the function looks smoother over the entire energy window, but shows some 
deviations from the target function in the vicinity of $E_c$.
It should be
noted that in the global update algorithm only two adjustable parameters
($a_4$ and $a_5$) are considered, whereas there are 30 free parameters 
widths in the local random scheme.  Fig. 3(b) shows the search for
a minimum $\Delta (a_5)$ at an optimum fixed $a_4$. For the given
expansion, one finds the optimal normalized expansion coefficients
$a^o_4 = 32.475$ and $a^o_5 = -10.600$ (dotted line), with a minimum
total error ($\approx 0.002 $) comparable to the local algorithm. The resulting 
globally adapted strength function is displayed in Fig. 3(c), and  in
Fig. 3(d) it is compared with the strength function of the local random 
update algorithm. While these strength functions share the qualitative 
similarity of a monotonic decay of the barrier widths as the boundaries
of the system are approached, obvious differences are that (i) the global 
update naturally results in a smoother strength function, (ii) 
this global strength function decays
more rapidly at the boundaries, and (iii) it vanishes exactly
at the boundaries. These last two features are due to the particular choice
of basis functions (polynomials), and the boundary conditions applied
to $s(x)$.  

Let us conclude the discussion of these two prototype adaptive 
design algorithms by applying them to less trivial target functions, such 
as filters with linearly and parabolically decaying transmission 
within given small energy windows. 
In Figs. 4(a) and (b) we compare the
results of the global and the local update algorithm. 
The decay from $T(E)=1$ (perfect transmission) to 0 (no transmission)
in the target functions (solid line) occurs in a small energy window 
$E \in
[0.42251 \; eV, 0.44251 \; eV]$. It can be seen that for these more complex
target responses, the local random update algorithm generally converges 
better than the global one, because the number of adaptive
parameters is much larger for the local method. 
The present implementation of
the global update algorithm is restricted to a two-dimensional
search, optimizing the coefficients $a_4$ and $a_5$. For the target step
function filter this method gives better results because here 
the coefficients $a_0$-$a_5$ decrease rapidly, and therefore higher-order
terms are not needed. In contrast, ``smoother" target filter functions require 
a larger number of coefficients to converge.
Therefore, in these cases the 
results for the global updates are not as good as for the guided random walk 
because of the restriction to a two-dimensional search. By increasing the 
number of coefficients in the global update algorithm a solution much closer
to the target filter function can be achieved.
Naturally, the local random update algorithm is much less sensitive to
these differences in the smoothness of the
target function, and typically converges fast
to an optimal symmetry-breaking configuration within 100 - 200 updates.\\

\section{Conclusions}
In summary, we have discussed numerical approaches to design arrays 
of nano-dielectrics to match desired target functions by intentionally
breaking translational symmetry. The global update algorithm is based
on an analytical expansion of the strength function of barrier
widths in terms of a basis, fixing the leading order expansion coefficients
by boundary conditions, and adjusting the remaining ones by numerical 
optimization. In contrast, the local random update approach uses each 
barrier width as an adjustable parameter. Sequential updates of
barrier widths are performed at random, and the updates are accepted 
when the resulting transmission profile matches better the target 
function than the previous one. Refinements of these prototype algorithms
are presently being investigated, including combinations of simultaneous
local and global updates and enforcement of lesser symmetries, such 
as point inversion. The broader aim behind these schemes is to develop
algorithms that are helpful in designing tools for emerging nanotechnologies.

\section{Acknowledgments}

We are grateful to T. Roscilde for useful
discussions, and acknowledge financial support by
DARPA, 
and the Department of Energy, Grant No. DE-FG03-01ER45908.

\onecolumngrid

\begin{figure}[h]
\caption {\label{fig1}}
\includegraphics[width=15cm]{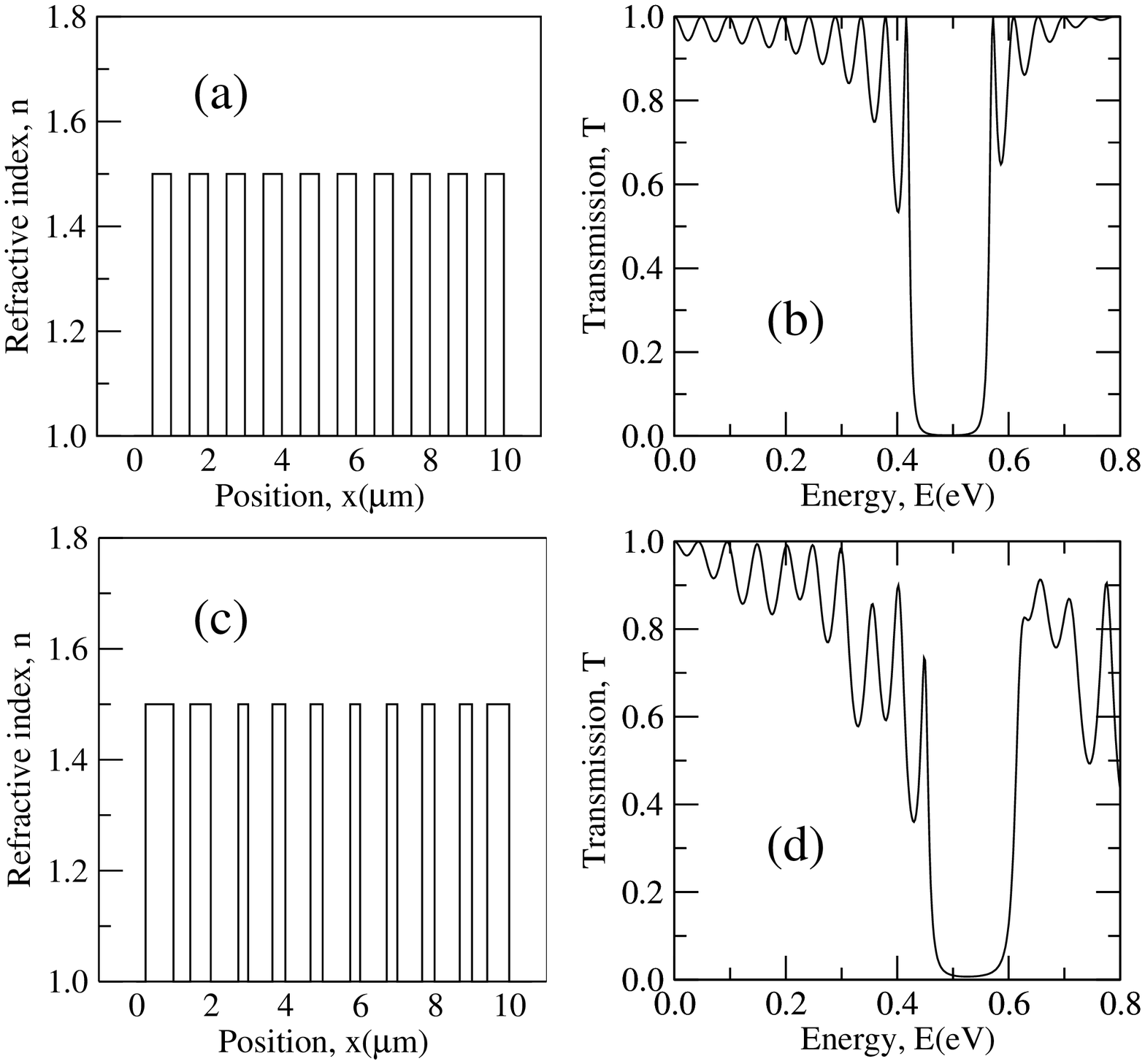}
\end{figure}

\newpage
\setlength{\baselineskip}{8mm}
\begin{center} {\large {\bf Table of Figures}} \end{center}
\textbf{FIG.~\ref{fig1}}.
Transmission as a function of energy in one-dimensional arrays 
of 10 optical wells and barriers with alternating refractive indices
$n_1 = 1.0$ and $n_2 = 1.5$, respectively. (a) and (b) Symmetric array with 
barrier widths of $0.5 \; \mu m$. (c) and (d) Slightly randomized
array with broken translational symmetry.
The duty cycle is fixed at $1 \; \mu m$, whereas the widths are generated 
from a uniform random distribution function, centered at $0.5 \; \mu m$.\\

\begin{figure}[h]
\caption{\label{fig2}}
\includegraphics[width=15cm]{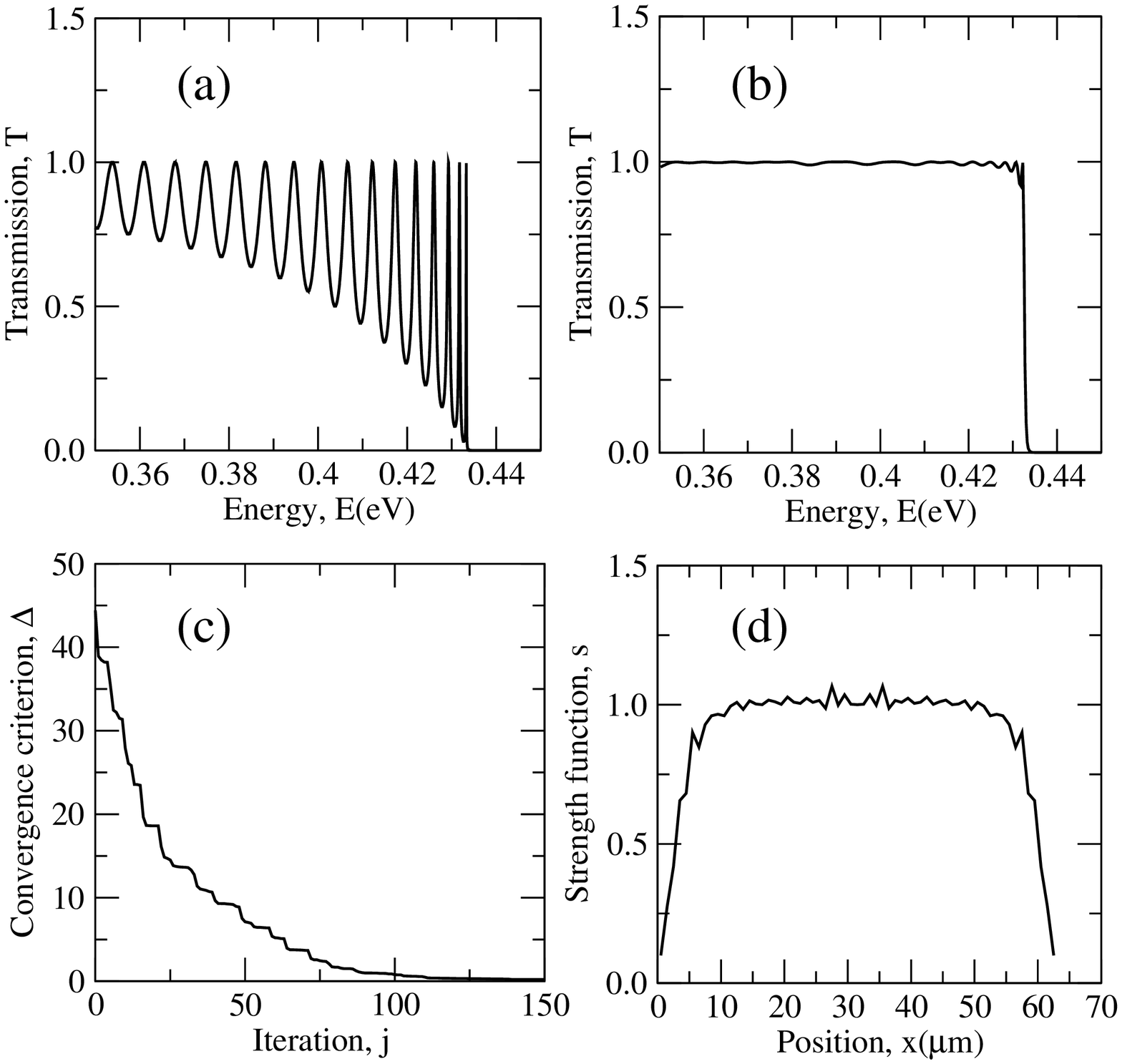}
\end{figure}

\noindent
\textbf{FIG.~\ref{fig2}}.
Adaptive random updates on an array with 30+2+30 optical barrier pairs. 
(a) Transmission as a function of energy for the translationally
symmetric case.
(b) Transmission as a function of energy for broken spatial symmetry
to achieve a cut-off filter function.
(c) Deviation of T(E) from target function.
(d) Corresponding strength function. \\

\begin{figure}[h]
\caption{\label{fig3}}
\includegraphics[width=15cm]{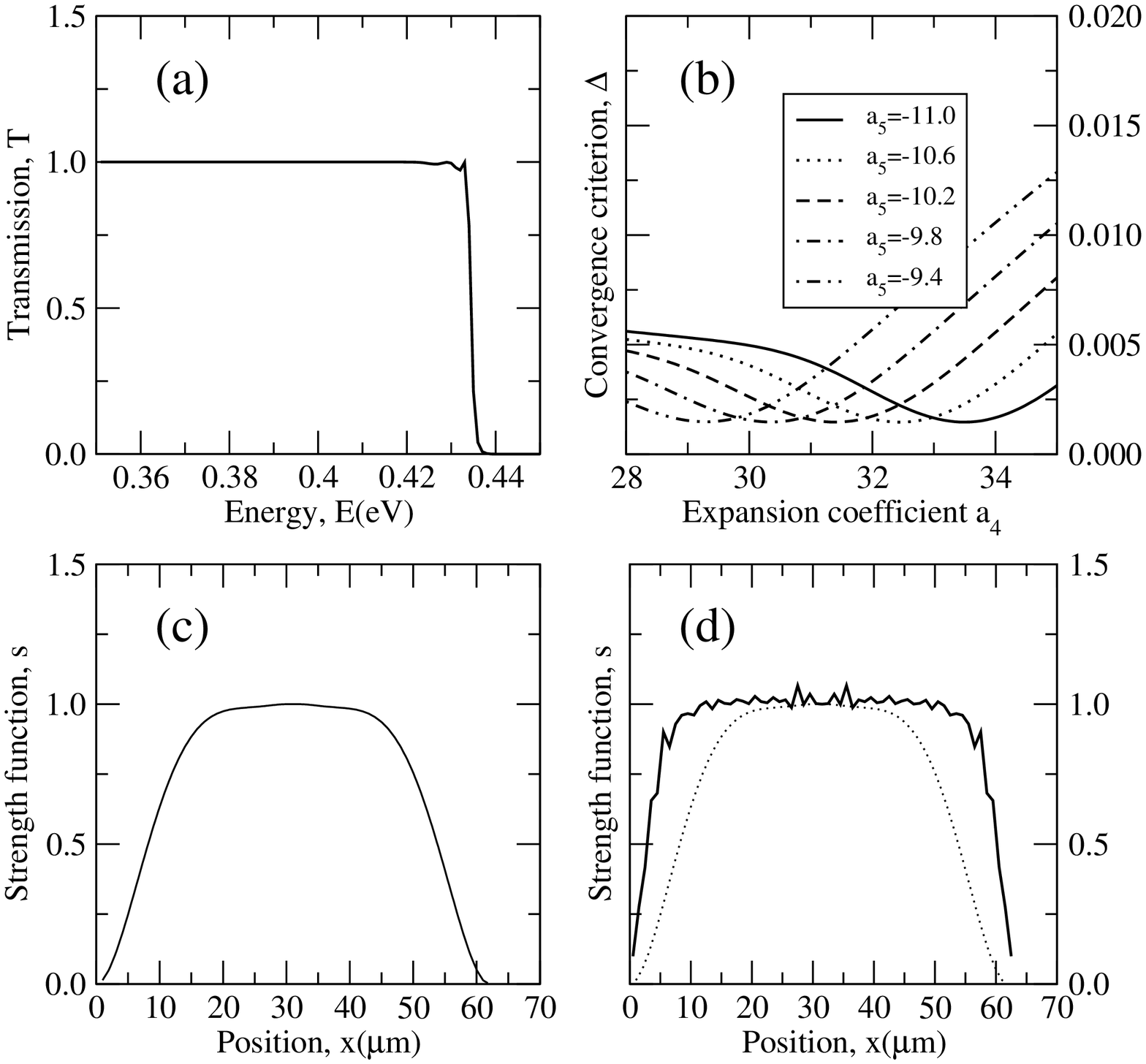}
\end{figure}

\noindent
\textbf{FIG.~\ref{fig3}}.
Adaptive global updates on an array with 30+2+30 optical barrier pairs. 
(a) Transmission as a function of energy
for broken spatial symmetry to achieve cut-off filter function.
(b) Optimization of $\Delta (a_5)$.
(c) Corresponding strength function. 
(d) Comparison of strength functions for random local (solid line)
and global (dashed line) update algorithms.\\

\begin{figure}[h]
\caption{\label{fig4}}
\includegraphics[width=15cm]{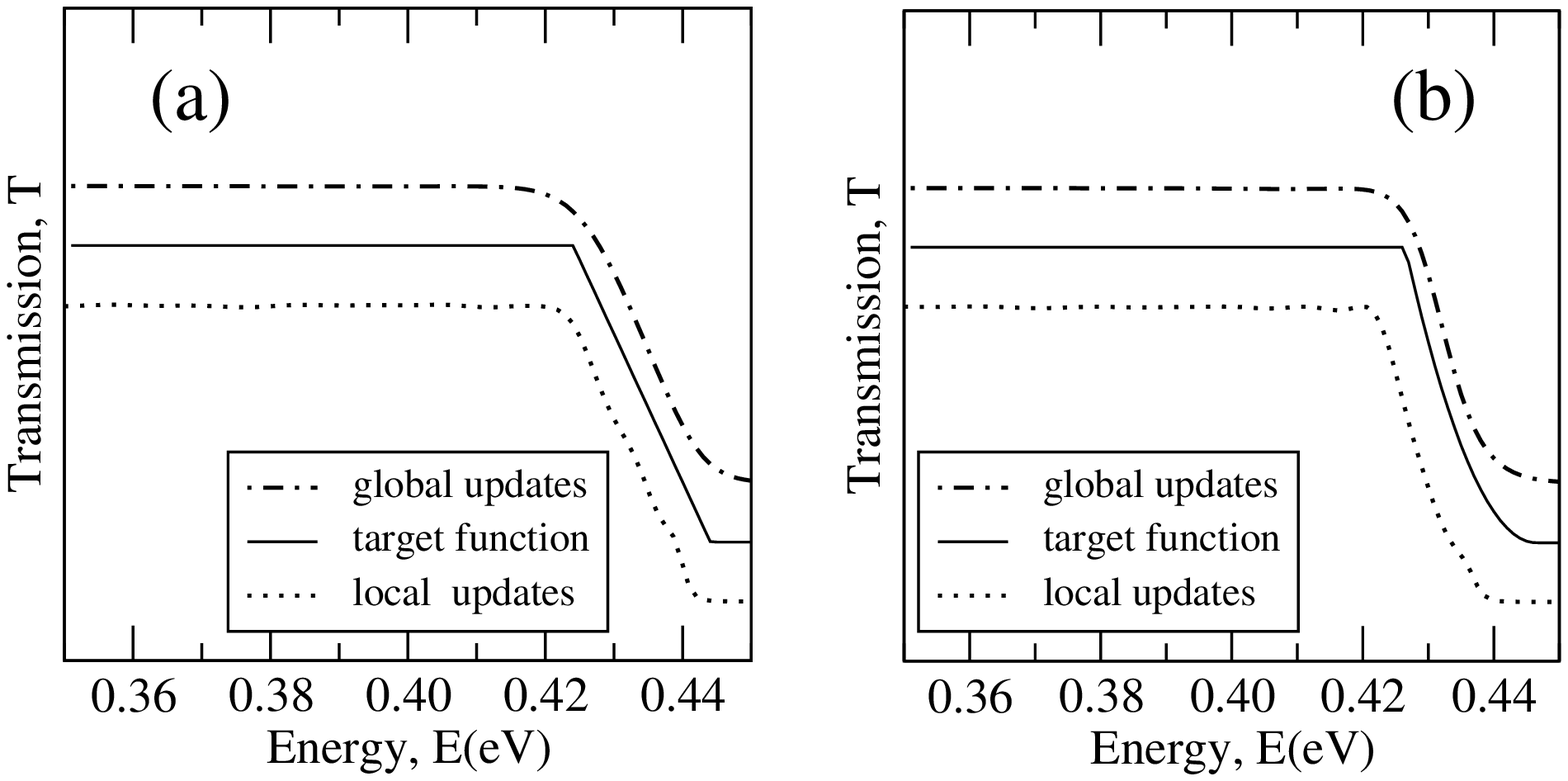}
\end{figure}

\noindent
\textbf{FIG.~\ref{fig4}}.
Comparison of the adaptive global update  
and local random update algorithms in a system with 15+2+15 optical barriers. 
(a) Linear target filter function. (b) Parabolic target filter function.
An off-set of $\pm$0.2 has been introduced to simplify the comparison.

\newpage

\newpage

\newpage

\newpage

\begin{references}
\bibitem{footnote1} The total number of dielectric
layers is 124. The widths of the 4
layers in the
center of the system are fixed at their initial values $b_0$, as discussed 
in the 
text. Since inversion symmetry about the array center is enforced in 
this example,
there remain 60 adjustable widths. Moreover, there is an additional
constraint that each pair of barriers 
satisfies $w_1(x) + w_2(x) =d$, where the $w_i$'s are 
the respective widths
of the on- and off-phase ($n_1$ and $n_2$), and $d$ is the overall 
constant width of 
each dielectric pair.
These constraints lead to a reduction of the number of free
parameters to 30.
\end{references}
\end{document}